\documentclass[nofootinbib,longbibliography,floatfix,superscriptaddress,11pt,tightenlines]{revtex4-1}
\usepackage[english]{babel}
\usepackage{amsmath,amssymb,amsbsy,amstext,amsthm,booktabs,simplewick,exscale,relsize,slashed,graphicx,amsfonts,upgreek}
\usepackage{multirow,dcolumn,bm,enumerate,url,hyperref}
\usepackage[usenames, dvipsnames]{color}

\newcommand{\gev}{\text{GeV}}
\newcommand{\tev}{\text{TeV}}

\begin{document}

\title{Multiscatter stellar capture of dark matter}

\author{Joseph Bramante}
\affiliation{Department of Physics, University of Notre Dame, 225 Nieuwland Hall, Notre Dame, Indiana 46556, USA}
\affiliation{Perimeter Institute for Theoretical Physics, 31 Caroline St. N, Waterloo, Ontario, Canada, N2L 2Y5}
\author{Antonio Delgado}
\affiliation{Department of Physics, University of Notre Dame, 225 Nieuwland Hall, Notre Dame, Indiana 46556, USA}
\author{Adam Martin}
\affiliation{Department of Physics, University of Notre Dame, 225 Nieuwland Hall, Notre Dame, Indiana 46556, USA}

\begin{abstract}
Dark matter may be discovered through its capture in stars and subsequent annihilation. It is usually assumed that dark matter is captured after a single scattering event in the star, however this assumption breaks down for heavy dark matter, which requires multiple collisions with the star to lose enough kinetic energy to become captured. We analytically compute how multiple scatters alter the capture rate of dark matter and identify the parameter space where the effect is largest. Using these results, we then show how multiscatter capture of dark matter on compact stars can be used to probe heavy $m_X \gg \tev$ dark matter with remarkably small dark matter-nucleon scattering cross-sections. As one example, it is demonstrated how measuring the temperature of old neutron stars in the Milky Way's center provides sensitivity to high mass dark matter with dark matter-nucleon scattering cross-sections smaller than the xenon direct detection neutrino floor.
\end{abstract}

\maketitle

\section{Introduction}
\label{sec:intro}

The nature of dark matter remains an outstanding mystery of our cosmos. Terrestrial direct detection experiments have become exceptionally sensitive to dark matter in the mass range ${\rm GeV} - {\rm 10~ TeV}$. While there are proposals for probing lighter dark matter, finding heavy dark matter, which has a lower particle flux through terrestrial detectors, presents a special challenge. Compact stars, which have a much larger fiducial mass than terrestrial detectors, provide an alternative means to probe dark matter.  Specifically, pairs of dark matter particles captured via interactions with the star can annihilate, leaving a distinct thermal trace.

Prior studies of dark matter's accumulation in stars have considered the case that dark matter capture occurs after dark matter scatters once off a stellar constituent ($e.g.$ nucleus, nucleon, electron). This is appropriate when the scattering cross section between dark matter and the constituent is small, leading to a mean path length that is large compared to the size of the star, so that at most one scatter is expected \cite{Press:1985ug,Gould:1987ir}. In this paper, we consider the case where the single scatter approximation breaks down and the dark matter is predominantly captured by scattering multiple times. We derive equations suitable for computing multiscatter capture of dark matter in stars, and as one application, show that observations of neutron stars in our galaxy would be sensitive to super-PeV mass dark matter that annihilates to Standard Model (SM) degrees of freedom, for dark matter-nucleon scattering cross-sections smaller than the xenon direct detection atmospheric neutrino floor.

To become captured while transiting through a star, dark matter must slow to below the stellar escape speed by recoiling against stellar constituents. During a single transit through the star, if the number of such interactions exceeds unity,
\begin{align}
N \approx n \sigma R \geq 1, 
\label{eq:ndim}
\end{align}
dark matter will be slowed (and possibly captured) by multiple scatters. Here $n$ is the number density of stellar constituents,  $R$ is the radius of the star, and $\sigma$ is the cross-section for dark matter to scatter off a stellar constituent. In white dwarfs, $\sigma$ is typically the cross-section for scattering off nuclei ($\sigma_{NX}$), while in neutron stars $\sigma$ is typically the cross-section for scattering off nucleons ($\sigma_{nX}$). One might also consider dark matter which predominantly scatters with electrons, in which case $\sigma$ would be the dark matter-electron cross-section. Often the stellar mass is related to the number of scattering sites by $M \simeq m\, N_n$, with $m$ the mass of a scattering site and $N_n$ the number of scattering sites per star. Keeping the stellar mass (or, equivalently, $N_n$) fixed while varying the star's size, Eq.~\eqref{eq:ndim} implies that the typical number of dark matter scatters inside a star scales as
\begin{align}
N \propto \frac{N_{n}\,  \sigma R}{\frac{4}{3} \pi R^3} \sim \frac{N_{n} \sigma}{R^2}.
\end{align}
As explored hereafter, this means that multiscatter capture is particularly relevant for dark matter accumulating in compact stars, $i.e.$ white dwarfs and neutron stars. Specifically, fixing $\sigma$ and comparing our Sun with an equivalent mass white dwarf ($R \sim 10^{-2}~ R_{\rm sun}$) or neutron star ($R \sim 10^{-5}~ R_{\rm sun}$), the smaller size of the compact stars leads to a $10^4$ enhancement in the average number of scatters for white dwarfs relative to the Sun, and a $10^{10}$ relative enhancement for neutron stars. 

While multiscatter can occur for dark matter of any mass, multiscatter capture is most important for heavy dark matter. This is primarily for two reasons. First, in order to be captured, the dark matter must lose a sufficient amount of its energy through collisions with scattering sites in the star. The fraction of the dark matter's energy lost in each collision depends on the scattering angle, but is proportional to the constituent mass $m$ divided by the dark matter mass $m_X$ in the limit that $m_X \gg m$. Therefore, heavier dark matter loses less energy per scatter, making gravitational capture after a single scatter less likely and multiscatter capture more important. Second, the range of dark matter-nucleon cross-sections for which heavy (PeV-EeV) dark matter capture in neutron stars proceeds predominantly through multiscatter energy losses, happens to coincide with dark matter-nucleon cross-sections just beyond the reach of next-generation direct detection experiments. Furthermore, it will be demonstrated in Section \ref{sec:results} that PeV-EeV mass dark matter can be captured by multiple ($\sim 10 - 10^3$ times) scatters in neutron stars even for the dark matter-nucleon cross-sections below the xenon direct detection ``neutrino floor," $\sigma_{\rm nX} \sim 10^{-45}~{\rm cm^2} (m_X/{\rm PeV})$ \cite{Billard:2013qya}. For these reasons, a primary focus of this paper will be dark matter with mass $m_{\chi} \gg \tev$. 

The dark matter masses just mentioned are well above the canonical WIMP mass scale of about 100 GeV. Dark matter with a weak scale mass has received deserved attention in the past decade because it can reproduce the observed dark matter abundance as a thermal relic. Considerable experimental efforts have bounded the nucleon scattering cross-section for weak-scale mass ($m_X \sim 100~{\rm GeV}$) dark matter to $\sigma_{nX} \lesssim 10^{-46}~{\rm cm^2}$, Ref.~\cite{Aprile:2012nq,Akerib:2016vxi,Tan:2016zwf}.  On the other hand, it has been shown that if one deviates from the minimal cosmological scenario, dark matter models with heavier masses $m_X \sim {\rm TeV-EeV}$ are predicted, $e.g.$ \cite{Kane:2011ih,Davoudiasl:2015vba,Randall:2015xza,Berlin:2016vnh,Bramante:2017obj}, either as a result of extra sources of entropy that dilute the thermal overabundance or because dark matter is very weakly coupled to the SM and it never thermalizes. 
As weak-scale mass dark matter has become increasingly constrained, the prospect of very heavy dark matter, which can still have a nearly ``weak" scale cross-section with nucleons ($\sigma \sim 10^{-40}~{\rm cm^2}$) deserves more attention. However, as a consequence of reduced dark matter flux, direct detection experiments have sensitivities that drop off with $1/m_X$ at high masses, and new methods to probe heavy dark matter are necessary. As we will show, neutron stars in our galaxy are powerful probes of heavy, weakly interacting dark matter.

Some prior work has considered multiscatter dark matter capture in the Earth and Sun \cite{Gould:1991va,Albuquerque:2000rk,Mack:2007xj}, where the gravitational potential of the capturing body, nuclear coherence, and relativistic effects could be reasonably neglected. Hereafter we treat single and multiple scatter capture rates and provide an equation valid for capture in the limit that the escape velocity of the capturing body greatly exceeds dark matter's halo velocity. The organization of the rest of this paper is as follows: in Section \ref{sec:simpeq}, we present our main points and the parametric dependence of multiscatter dark matter capture in compact stars. A detailed derivation of multiscatter capture is given in Section \ref{sec:detail}. Using the derived multiscatter capture formulae, Section \ref{sec:results} finds prospects for old neutron stars near the galactic center to constrain heavy dark matter that annihilates to Standard Model particles. In Section \ref{sec:conclusions}, we conclude.

\section{Parametrics of multiscatter capture}
\label{sec:simpeq}

In order to calculate the parametric dependence of multiscatter capture, we are going to first examine the dark matter single-scatter capture rate, and then investigate how the rate changes when one accounts for more than one collision. We will find that, for heavy enough dark matter, the mass capture rate of dark matter on compact stars depends linearly on $\sigma$ and inversely on $m_X$.  This $\sim \sigma/m_X$ scaling of the mass capture rate arises for heavier dark matter, because more scatters (which scale up with $\sigma$) are needed for heavier particles to be captured by the star.

Dark matter capture in a star depends upon the flux $F$ of dark matter through the star and the probability $\Omega$ that collision(s) with the star will deplete the dark matter's energy enough that it becomes gravitationally bound. The flux in turn depends upon the number density of dark matter in the halo $\left( n_X=\frac{\rho_X}{m_X} \right)$, the relative motion of the star with respect to the dark matter halo ($v_{star}$), the distribution of dark matter speeds in the dark matter halo, and the escape speed of the dark matter halo ($v^{halo}_{esc}$). The probability to capture ($\Omega$) depends on the speed of the dark matter, set by the initial speed plus the amount of speed it has gained falling into the star's gravitational well. Additionally, the probability depends on the density of scattering sites in the star ($n_T$), the cross section of dark matter to scatter off scattering sites ($\sigma$), and the fraction of scattering phase space where sufficient energy is lost. Both the velocity gained by falling into the star and the number density are, in principle, functions of where inside the star the collision occurs. Combining the flux and capture probability yields a differential capture rate, which must be integrated over dark matter initial velocities and trajectories through the star. Schematically, the differential capture rate is
\begin{align}
\frac{d\, C}{dV\, d^3u} = dF(n_X, u, v_{star}, v^{halo}_{esc})~\Omega(n_T(r), w(r), \sigma, m, m_X),
\end{align}
where $u$ is the dark matter velocity far from the star (the halo velocity) and $w^2(r) = u^2 + v^2_{esc}(r)$ is the speed of the dark matter after it has fallen to a distance $r$ from the star's center (either inside or outside of the star).

To focus on the parametrics of dark matter capture, for simplicity we assume no motion of the star relative to the dark matter thermal distribution in the halo ($v_{star} \to 0$) and an infinite escape speed for the dark matter halo ($v^{halo}_{esc} \to \infty$). 
We also fix the escape speed of dark matter in the star to the escape speed at the star's surface ($v_{esc}(r) = v_{esc}(R)$), and for the moment omit general relativistic and nuclear physics corrections. With these provisos, a constant-density star in the rest frame of the dark matter halo with stellar escape velocity $v_{\rm esc}^2 \sim 2 G M /R$ has a single-scatter dark matter capture rate derived in Appendix \ref{app:single}
\begin{align}
C_{1} = \sqrt{24 \pi} G  \frac{\rho_X}{m_X} M R  \frac{1}{\bar{v}} ~{\rm Min}\left[1, \frac{\sigma}{\sigma_{\rm sat}}\right] \left( 1-\frac{1-e^{-A^2}}{A^2} \right). \label{eq:singlescatter}
\end{align} 
Note that the capture rate scales with dark matter density $\rho_X$ and inversely with the dark matter halo velocity $\bar v$. Here, $G$ is Newton's constant, $M$ is the mass of the star,  $\sigma$ is dark matter's cross-section with a stellar constituent (nucleus, nucleon, electron). The exponential factor $A^2 \equiv \frac{3}{2} \frac{v_{\rm esc}^2}{\bar{v}^2} \beta_-$, where $\beta_\pm \equiv 4 m_X  m/(m_X \pm m)^2$ and $m$ is the mass of the particle (nucleus, nucleon, electron) dark matter scatters against. Increasing the cross-section past a certain threshold will guarantee that most transiting dark matter scatters with the star at least once, though it may not lose enough energy to be captured. This threshold cross-section is customarily defined as $\sigma_{\rm sat} = \pi R^2 /N_n$, where $N_n$ is the number of scattering sites, and the ``Min" function evaluates to unity once at least one capture is probable. The parenthetical term in Eq.~\eqref{eq:singlescatter} takes into account dark matter that scatters but does not lose sufficient energy to be gravitationally captured.

To better understand the origin of the parenthetical piece of Eq.~(\ref{eq:singlescatter}), let us examine the energetics of gravitational capture. To be captured after a single collision, the energy lost by the dark matter must be greater than its initial kinetic energy in the galactic halo. The energy loss is proportional to the reduced mass of the dark matter - constituent system, $\mu_n$ and the speed of the dark matter at the collision site. In the limit that the star's escape velocity is much greater than the halo velocity ($w = \sqrt{u^2+ v^2_{esc}}  \simeq v_{esc}$) the capture requirement is
\begin{align}
\Delta E \simeq 2\, \frac{\mu^2_n}{m}\, v^2_{esc}\, z \ge \frac 1 2 m_X\, u^2, 
\label{eq:cap}
\end{align}
where $z$ is a kinematic variable $\in [0,1]$ related to the scattering angle. Assuming dark matter is much heavier than the stellar constituents and turning the above requirement above into a condition on $u$,
\begin{align}
u < u_{max} = \sqrt{\beta_+\, z}\, v_{esc}.
\end{align}
In the full capture treatment (Appendix \ref{app:single}), for dark matter with Boltzmann distributed velocities from $0$ to $u_{max}$ and scattering angles $z \in [0,1]$, we consider kinematic phase space where dark matter is moving slowly enough to be captured after a single collision. The limit of this phase space is set by $u_{max}$, which is evident in the form of the $A^2$ exponential factor in Eq.~\eqref{eq:singlescatter}. Note that when $m_X \gg m$, a limit that will be appropriate throughout this paper, $\beta_{\pm}$ both reduce to $4\, m/m_X$. 

The origin and form $A^2$ term are important because $A^2$ governs the dependence of $C_1$ on the dark matter mass.
When $A^2$ is large, corresponding to a maximum capture speed much larger than than average dark matter speed,  the parenthetical term in Eq.~(\ref{eq:singlescatter}) evaluates to 1, and the sole dark matter mass dependence lies in the number density $\frac{\rho_X}{m_X}$. In this case, the single scatter capture rate scales as
\begin{align}
C_1 \propto \frac{\sigma}{m_X}, ~~~~~(A^2 \gg 1)
\label{eq:Asq1}
\end{align} 
implying a mass capture rate $m_X C_1 \propto \sigma$ that is independent of the dark matter mass. However, if $A^2$ is small, implying a maximum capture speed less than a typical dark matter halo velocity $\bar{v}$, we can expand the entire parenthetical expression in Eq.~\eqref{eq:singlescatter}, and find that the capture rate scales as
\begin{align}
C_1 \propto \frac{\rho_X}{m_X}\,\, \sigma \,A^2 \propto \frac{\sigma}{m^2_X},~~~~~(A^2 \ll 1)
\label{eq:Asq2}
\end{align}
implying a mass capture rate scaling 
$
m_X C_1 \propto \sigma/m_X
$
that depends inversely on the dark matter mass.

To see where the mass capture rate transitions from being constant to being $m_X$-dependent in compact stars, we can insert appropriate values for $v_{esc}$. For a solar mass white dwarf $v_{esc}\,c \sim 2 \times 10^3\, \text{km/s}$, while a solar mass neutron star has $v_{esc}\,c \sim 2\times 10^5\, \text{km/s}$; both of these escape speeds are far greater than the average dark matter halo speed $\bar v c\sim 220 \,\text{km/s}$, therefore $A^2$ will only be less than one if the dark matter is much heavier than $m$. Specifically, taking $A^2 = 1$ to be the transition value, and solving for $m_X$, we find the transition occurs at $m_X \sim \tev$ in a solar mass white dwarf (assuming scattering off of carbon) and $m_X \sim \text{PeV}$ for a solar mass neutron star (assuming scattering off a neutron). 

To see how the parametric dependence of Eq.~\eqref{eq:singlescatter} changes in the case of multiple scatters, let us revisit the energetics of gravitational capture. For the moment, let us assume that dark matter participates in $N \ge 1$ collisions during its transit of the star and that each collision results in an average energy loss
\begin{align}
\Delta E_i = \frac{\beta_+ E_i}{2}.
\end{align}
If the dark matter initially entered the star with energy $E_0$, the energy after $N$ `average' collisions is
\begin{align}
E_N = E_0 \left( 1 - \frac{\beta_+}{2} \right)^N, 
\end{align}
or a net energy deposit of $\Delta E_N = E_0 - E_N$. Assuming, as in the single scatter case, that the initial dark matter kinetic energy is $E_0 \sim 1/2\, m_X\, v^2_{esc}$ and plugging $\Delta E_N$ into the capture condition Eq.~(\ref{eq:cap}), we can solve for the maximum halo velocity $u$ that can be captured
\begin{align}
u \le v_{esc}\, \Big( 1 - \left( 1 - \frac{\beta_+}{2} \right)^N\Big)^{1/2}
\label{eq:simplemulti}
\end{align}
In the limit that $m_X \gg m$ and $\beta_+ \rightarrow 4 m/m_X$, the leading order term in the binomial expansion of the right side of Eq.~\eqref{eq:simplemulti} approximates the full expression.  In that limit, the maximum allowed velocity simplifies to
\begin{align}
u \le \sqrt{\frac{N\, \beta_+}{2}}\, v_{esc} \cong \sqrt{\frac{2\, N\, m}{m_X}}\, v_{esc}
\end{align}
up to corrections of $\mathcal O \left( \frac{(N\,m)^2}{m^2_X} \right)$. As we will show in more detail in the next section, in the limit of $v_{esc} \gg \bar v$ the probability to capture after $N $ scatters can be expressed in a form very similar to \eqref{eq:singlescatter} but with $A^2$ -- the factor in the exponential -- modified to
\begin{align}
A^2_N \equiv \frac{3\, v^2_{esc}}{\bar v^2}\frac{N\, m}{m_X}.
\label{eq:newlim}
\end{align}
As discussed following Eq.~(\ref{eq:singlescatter}), if this exponential factor is large then the $m_X$-dependence in the capture rate from the $A^2$ term is suppressed. Meanwhile, if the factor is small, the exponential can be approximated by an expansion, resulting in a capture rate $\propto  n_X\, A^2  \propto \sigma/m^2_X$. Comparing Eq.~\eqref{eq:singlescatter} to Eq.~\eqref{eq:newlim}, we see that multiple scattering has added a factor of $N$ to the $A^2$ term. The $N$ dependence in the numerator of Eq.~(\ref{eq:newlim}) means that for $N \gg 1$,  the dark matter mass needs to be larger (for a given $v_{esc}, \bar v$ and $m$) before the exponential factor becomes small. Stated another way, if the dark matter scatters $N$ times, the capture rate will behave as $C_N \sim \sigma/m_X$ out to masses $N$ times higher than if dark matter only scatters once. 

Note that this discussion has involved only the energetics of slowing down a heavy dark matter particle to beneath a star's escape speed and not whether the dark matter interacts with stellar constituents strongly enough to participate in multiple scatters in the first place. Following from Eq.~(\ref{eq:ndim}), the likelihood to participate in multiple scatters roughly depends on the path length of the dark matter $1/n \sigma $ compared to the size of the star. We will flesh out this dependence in the next section.

\section{Multiscatter capture}
\label{sec:detail}

Having examined the parametric scaling of multiscatter capture in the previous section, in this section we derive the multiscatter dark matter capture rate. Our notation follows that of \cite{Gould:1991va}, which considered capture by the Earth's iron core, where the acceleration of incoming dark matter due to Earth's gravity, and -- more broadly -- general relativistic effects, could be neglected. In the large $N$ limit, the treatment presented here also allows for more efficient computation of the multiscatter capture rate, by obviating the $N$-fold kinematic phase-space integral in \cite{Gould:1991va}.

For multiscatter capture it is convenient to define the optical depth $\tau = \frac{3 \sigma}{2\sigma_{\rm sat}}, \sigma_{sat} = \frac{\pi R^2}{N_n}$, the average number of times a dark matter particle with dark matter - nuclear cross section $\sigma$ will scatter when traversing the star.\footnote{To understand the $\frac{3}{2}$ factor in the optical depth, observe that the cross section for which $1$ scatter occurs over a distance of $2R$, (where $R$ is the radius of the star) is
\begin{align}
1 = n\, \sigma\, (2R) &= \frac{N_n}{(4/3)\pi R^3}\sigma\, (2R) = \frac{3\, N_n}{2\pi\, R^2}\sigma \\ \nonumber
& \rightarrow \sigma = \frac{2}{3} \left(\frac{\pi R^2}{N_n} \right) = \frac{2}{3} \sigma_{\rm sat},
\end{align}
The optical depth is normalized so that $\tau =1$ when dark matter typically scatters once as it passes through the star.} The probability for dark matter with optical depth $\tau$ to participate in $N$ actual scatters is given by $\text{Poisson}(\tau, N)$. However, this expression can be improved to incorporate all incidence angles of dark matter. Defining $y$ as the cosine of the incidence angle of dark matter entering the star, the full probability is
\begin{align}
p_N(\tau) = 2 \int_0^1 dy~\frac{y e^{-y \tau} \left(y \tau \right)^N}{N!}.
\label{eq:poisson}
\end{align}
While it incorporates all incidence angles, this expression still makes the assumption that the dark matter takes a straight path through the star. In practice, the straight path assumption will produce conservative bounds on dark matter capture, marginally under-predicting the capture rate. 

Incorporating the likelihood for dark matter to participate in $N$ scatters, the differential dark matter capture rate after exactly $N$ scatters looks similar to the single scatter formula (see Appendix \ref{app:single}), with the probability to capture after $N$ scatters $g_{N}(w)$ adjusted to take into account the kinematics of $N$ collisions and replacing $\frac{\sigma}{\sigma_{sat}} \rightarrow p_{N}(\tau)$\footnote{The multi-scatter capture rate \eqref{eq:dCn} can be obtained by setting $n(r) = \frac{N_n}{\frac{4}{3} \pi R^3}$ in Eq.~\eqref{eq:g1dCsingle}, integrating $r$ from $0$ to $R$ and making the substitutions $g_1(w) \rightarrow g_{N}(w)$ and $\frac{\sigma}{\sigma_{sat}} \rightarrow p_{N}(\tau)$.},
\begin{align}
C_N & = \pi R^2 \, p_{N}(\tau) \int_{0}^{\infty}  ~ f(u) \frac{du}{u}~ w^2  g_N(w).
\label{eq:dCn}
\end{align}
The velocity distribution $f(u)$ of dark matter particles in the galactic halo is given in Eq.~\eqref{eq:mboltz}. 
In writing the velocity distribution as $f(u)$ we have retained the assumptions from the single capture case that the escape velocity of the dark matter halo is infinite and the velocity of the star relative to the dark matter is zero. We have also maintained that the density of the star is uniform and ignored the radial dependence of the escape velocity.\footnote{To estimate how much the constant density assumption alters the neutron star capture rate, consider an approximate neutron star density profile (ADP) $\rho_{NS}^{\rm ADP} (r) = 2.6 \times 10^{38} ~{\rm GeV/cm^3} \left( \frac{10~{\rm km}}{r} \right) $, which matches a 1.5 $M_{\odot}$, $R=$ 10 km neutron star. This can be compared to a constant density (CD) profile, such a neutron star would have $\rho_{NS}^{\rm CD} \simeq 4 \times 10^{38} ~{\rm GeV/cm^3}$. We can calculate the integrated optical depth $d \tau_{i} = n(r) \sigma_{nX} d \ell$, where $\ell$ is the path of the dark matter particle. Calculating this integrated optical depth for a dark matter particle that passes within a kilometer of the center of the neutron star, we find that for the constant density and approximate density profile cases, for trajectories passing deep within the neutron star, the optical depth can increase by up to fifty percent. This would somewhat aid capture in the multiscatter regime. Therefore, the bounds derived in this paper are somewhat conservative.}

It is convenient to shift the integral to $w$, where $w^2 =u^2 + v_{\rm esc}^2$. The capture rate for $N$ scatters then becomes
\begin{align}
C_N =  \pi\, R^2 \, p_{N}(\tau) \int_{v_e}^{\infty}\, dw\, \frac{f(u)}{u^2}\, w^3\, g_N(w),
\label{eq:dCn2}
\end{align}
and the total capture rate is the sum over all $N$ of the individual $C_N$
\begin{align}
C_{\rm tot} = \sum_{N=1}^{\infty} C_N.
\label{eq:CNsum}
\end{align} 
In actual computations, the sum in Eq.~(\ref{eq:CNsum}) will be cut off at some finite $N_{max}$ where $p_{N_{max}}(\tau) \approx 0$.

Finally, we need to evaluate $g_N(w)$, the probability that the speed of the dark matter after $N$ collisions drops below the escape velocity. This probability, which we analyzed dimensionally in Section \ref{sec:simpeq}, depends solely on dark matter's initial velocity, the amount of energy lost in each scatter, and the escape velocity of the star. For dark matter with initial kinetic energy at the star's surface $E_0 = m_X w^2/2$, the energy lost in a single scattering event is given by $\Delta E = z \beta_+ E_0 $, where $z$ is related to the scattering angle, $z \in [0,1] $, and we again note that $\beta_+ \equiv 4 m_X  m/(m_X + m)^2$.  Iterating for $N$ scatters, the dark matter energy and velocity decrease to
\begin{align}
E_N = \prod_{i=1}^N\, (1-z_i\, \beta_+)\, E_0,~~~~ v_N = \prod_{i=1}^N\, (1-z_i\, \beta_+)^{1/2}\, w.
\end{align}
If the velocity after $N$ scatters is less than the escape velocity, the dark matter is captured. Phrased as a condition on the initial velocities $w$ that we are integrating over, the capture probability is
\begin{align}
g_N(w) = \int_{0}^1\,dz_1\int_{0}^1\,dz_2 \cdots\int_{0}^1\,dz_N\, \Theta\Big(v_{esc}\prod_{i=1}^N(1-z_i\,\beta_+)^{-1/2} - w\Big),
\label{eq:gnfull}
\end{align}
where the $dz_i$ integrals sum over all possible scattering trajectories (angles) at each step. This condition requires an integral for every scatter, and becomes computationally cumbersome to evaluate for large $N$. Therefore, as a further approximation, let us replace the $z_i$ with their average value. Provided the differential dark matter-nuclear cross section is independent of scattering angle -- valid in most scenarios of spin-independent elastic scattering -- $\langle z_i \rangle \approx 1/2$ and $g_N(z)$ simplifies\footnote{We have checked numerically that for $N \gtrsim 5$, the approximate expression in Eq.~(\ref{eq:uNrel}) matches the full expression Eq.~(\ref{eq:gnfull}) to within less than a percent for the applications presented in Section \ref{sec:results}.},
\begin{align}
g_N(w) = \Theta\Big(v_{esc}(1-\langle z_i \rangle \beta_+)^{-N/2} - w\Big).
\label{eq:uNrel}
\end{align} 
As in the single scatter case, the capture probability restricts the range of dark matter velocities that allow for capture. To illustrate the relationship between dark matter's halo speed and the number of scatters it takes to slow down to below the star's escape speed, we recast Eq.~(\ref{eq:uNrel}) as contours in $u-N$ space in Fig.~\ref{fig:uNfig} below, for typical neutron star and white dwarf parameters (see caption).
\begin{figure}[h!]
\centering
\includegraphics[width=0.47\textwidth]{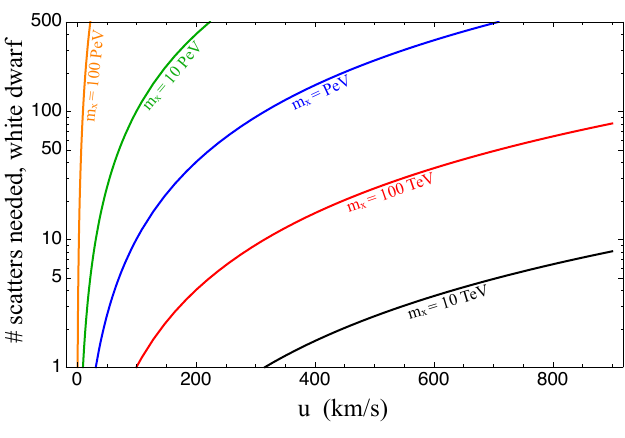}
\includegraphics[width=0.47\textwidth]{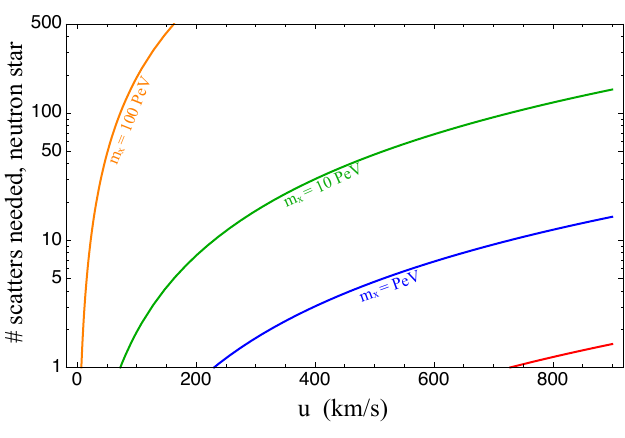}
\caption{Number of scatters needed to capture dark matter as a function of dark matter's halo speed ($i.e.$ the speed at long distance from the star). The left plot shows the relation assuming a solar mass white dwarf made entirely of carbon ($m_N \sim 10\, \gev$) and with radius $R = 0.1\, R_{sun}$. The right plot shows the relation for a solar mass neutron star with radius $R = 10\, \text{km}$, which for the moment neglects relativistic corrections. The lines correspond to 10 TeV--100 PeV mass dark matter, as indicated.}
\label{fig:uNfig}
\end{figure}
The fact that dark matter with a given mass and speed requires more scatters to be captured in a white dwarf is due to the fact that the velocity at infinity ($u$) is a larger fraction of the star's escape speed than for a neutron star. This gives the impression that multiscatter is more important for dark matter capture in white dwarfs. However, the number of scatters needed to slow down to sub-escape velocities is not the only factor in the problem; capture also depends on whether the dark matter-nuclei cross section is large enough for the dark matter to interact scatter multiple times as it transits the star.  The strength of the dark matter - constituent interaction is encapsulated in the optical depth $\tau$ which, as we have seen, is proportional to $1/R^2$ and therefore much larger for neutron stars.

 Using the simplified form for $g_N(w)$, we can evaluate remaining integral in Eq.~(\ref{eq:dCn2}):
\begin{align}
C_N =  \pi\, R^2\, p_N(\tau) \frac{\sqrt 6\, n_X}{3 \sqrt{\pi}\bar v}\Big((2\,\bar v^2 + 3\, v^2_{esc}) - (2\, \bar v^2 + 3\, v^2_N)\exp{\Big(-\frac{3(v^2_N - v^2_{esc})}{2\,\bar v^2} }\Big)\Big),
\end{align}
with $v_N = v_{esc}(1 - \beta_+/2)^{-N/2}$. In the limit that $v_{esc} \gg \bar v$ and $m_X \gg m$, this becomes
\begin{align}
C_N = \sqrt{24\,\pi}\,p_N(\tau)\, G\, n_X\, M\, R\frac{1}{\bar v}\left( 1 - \left(1 + \frac{2 A^2_N\, \bar v^2}{3\, v^2_{esc}} \right)\, e^{-A^2_N} \right); \quad A^2_N = \frac{3\, v^2_{esc}\, N m}{\bar v^2\, m_X},
\label{eq:CNpart}
\end{align}
where the last expression follows the format of the single scatter capture equation Eq.~\eqref{eq:singlescatter}. Note that the reason $C_1$ according to this formula does not precisely match Eq.~\eqref{eq:singlescatter} is that we integrated over all possible energy loss fractions ($dz_1$) when deriving the latter, but assume average energy loss in the former. As expected, the capture rate for $N$ scatters has a similar form as the single capture rate, up to a factor of $N$ in the exponential factor $A^2_N$. Following the logic presented in Sec. \ref{sec:simpeq}, the factor of $N$ implies the $C_N \propto 1/m_X$ scaling persists out to higher $m_X$ than in the single scatter case. However, while the behavior of an individual $C_N$ is easy to see given $m_X, m$ and $v_{esc}$, the mass scaling of the full capture rate is more subtle as it involves the sum over all $C_N$, each weighted by $p_N(\tau)$.  \\

Having reviewed the general form of the multiple scatter capture rate, we can now apply it to white dwarfs and neutron stars. Each of these applications involves subtleties not present in Eq.~\eqref{eq:dCn2}.

White dwarfs are compact stars ($R \sim 10^4\, \text{km}, M \sim 10^{57}~{\rm GeV}$) that are supported by electron degeneracy pressure. Their suitability as potential laboratories to capture and thereby constrain various dark matter candidates has been studied previously in the single-scatter regime \cite{Bertone:2007ae,Kouvaris:2010jy,McCullough:2010ai,Bramante:2015cua,Graham:2015apa}. At the upper end of the mass range, white dwarfs are largely composed of carbon and oxygen, so $m = m_N \sim \mathcal O(10 ~{\rm GeV})$ in the capture equations above.  Dark matter possessing spin-independent ($e.g.$ scalar or vector current)  interactions with nuclei will scatter coherently off the nucleons within carbon/oxygen if the momentum exchange is low enough, while higher energy exchanges will be sensitive to the substructure of the nucleus and correspondingly suppressed. This loss of coherence is expressed by a form factor.
Including the form factor suppression, the multiscatter accumulation rate will be given by Eq.~\eqref{eq:CNsum} with the cross-section substitution
\begin{align}
\sigma \rightarrow \sigma^{\rm WD}_{NX} \simeq \sigma_{nX} \frac{m_{N}^4}{m_{n}^4} F^2(\langle E_{\rm R} \rangle) ,
\label{eq:snx}
\end{align}
where, in the case of scattering off carbon, the mass of the stellar constituent is $m_{N} \simeq 12\, m_n\, \simeq 11.1 ~{\rm GeV}$, and $F^2(\langle E_{\rm R} \rangle)$ is the Helm form factor evaluated at the average recoil energy $\langle E_{\rm R} \rangle$ \cite{Helm:1956zz}. The average recoil energy is defined as
\begin{align}
\left\langle E_{\rm R} \right\rangle \simeq \frac{\int_{0}^{E_{\rm R}^{\rm max}} ~dE_{\rm R}~ E_{\rm R}F^2(E_{\rm R})}{\int_{0}^{E_{\rm R}^{\rm max}}~dE_{\rm R}~  F^2(E_{\rm R})},
\end{align}
where we make the approximation that $v_{\rm esc}$ is much greater than the halo velocity and therefore $E_{\rm R}^{\rm max} \simeq 2 m_N v_{\rm esc}^2$. For recoil energies relevant for heavy dark matter scattering off carbon in a solar mass white dwarf ($v_{\rm esc} \simeq 0.01$, $\langle E_{\rm R}\rangle \simeq {\rm MeV}$), the form factor evaluates to $F^2(\langle E_{\rm R} \rangle) \sim 0.5$.

\begin{figure}[h!]
\includegraphics[scale=.75]{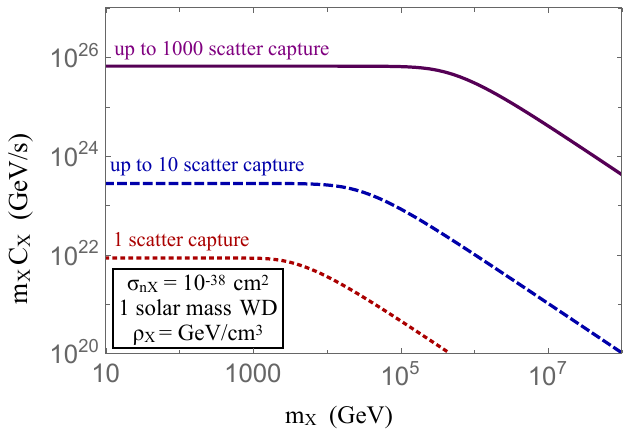}
\includegraphics[scale=.75]{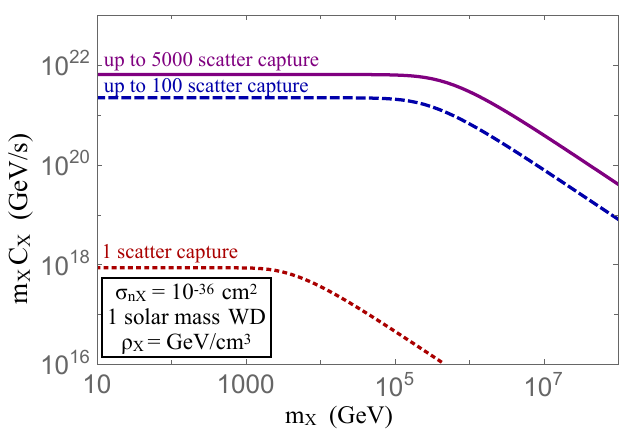}
\caption{Mass capture rate of dark matter on a constant density white dwarf, for a per-nucleon scattering cross-section of $\sigma_{nX} = 10^{-38}$ (left panel) and $10^{-36}$ cm$^2$ (right panel). Following Eq.~\eqref{eq:snx}, these per-nucleon cross sections translate to dark-matter carbon cross sections of $\sigma_{nX} \sim 10^{-34}$ and $\sim 10^{-32}$ cm$^2$. In both panels we have taken the target star to be a 1 solar mass white dwarf composed of carbon 12, with $R=10^4$ km, in a background dark matter density of $\rho_{X} =  {\rm GeV/cm^3}$ with halo velocity dispersion $\bar{v} \simeq 220~{\rm km/s}$.}
\label{fig:wdmc}
\end{figure}

In addition to affecting the overall scattering cross section, the form factor also impacts the weighting of different momentum exchanges (scattering angles) in each scatter, previously encapsulated in the variable $z_i$. Higher momentum exchanges are suppressed by the form factor as they correspond to reduced dark matter-nucleus scattering coherence. As a result, lower energy scatters -- where a smaller fraction  the dark matter's kinetic energy is deposited in each scatter  -- are more common. To account for this, we make the substitution $\left\langle z_i \right\rangle \approx \left\langle E_{\rm R} \right\rangle/E_{\rm R}^{\rm max}$ (instead of  $\left\langle z_i \right\rangle=\frac{1}{2}$) in Eq.~\eqref{eq:uNrel}. In particular, for the numerical results given in Figure \ref{fig:wdmc}, we approximate $\left\langle z_i \right\rangle =0.15$. In deriving $\left\langle z_i \right\rangle$, we have assumed that the relative velocity of the dark matter and nucleus remains constant (at $\sim v_{\rm esc}$) during the capture process. This assumption is valid so long as the dark matter halo velocity is much smaller than its velocity during capture $u \ll w \sim v_{\rm esc}$, implying that the speed of the dark matter remains approximately constant during capture. To understand this, note that as soon as the dark matter velocity decreases by an $\mathcal{O}(1)$ factor from $w \sim v_{\rm esc}$, its speed will be well below the escape velocity, since $u \ll v_{\rm esc}$.

The mass capture rates for heavy dark matter in a white dwarf, computed using both the single and multiple capture expressions and two different assumptions about the size of the dark matter-nucleon cross section are shown in Fig.~\ref{fig:wdmc}. The contours in Figs.~\ref{fig:wdmc} display the capture rate for up to $N \leq 1,10, 100...$ scatters, using Eq.~\eqref{eq:CNsum}. As the dark matter-nucleon cross-section increases, the difference in mass capture rate for  $N = 1$ versus $N \leq 1000$ scatters increases dramatically. This is a consequence of the fact that, as the dark matter-nucleon cross-section becomes large enough, most trajectories through the white dwarf will involve multiple scattering events and so the rate for capture after a single scatter more substantially under-predicts the total capture rate. We can also see that, as the number of scatters increases, the ``turnover mass" (the mass at which the capture rate diminishes) also increases. As explored in Section \ref{sec:simpeq}, this is because lighter dark matter requires fewer scatters to be captured, since the fractional energy loss of the dark matter per scatter is $\sim 2 m_{\rm N} /m _X$. 

The quoted per-nucleon scattering cross-sections in Fig.~\ref{fig:wdmc}, $\sigma_{nX} =10^{-34}$ and $10^{-36}~{\rm cm^2}$, which were chosen to be large enough so that multiple scatters are relevant, are typically excluded by direct detection searches for spin-independent DM-nucleon scattering \citep{Aprile:2012nq,Akerib:2016vxi,Tan:2016zwf}. One might consider whether white dwarfs could be used to constrain spin-dependent DM-nucleon interactions, which are less constrained by direct detection searches. Unfortunately, white dwarfs are composed of mainly spin-free nuclei ($e.g.$ carbon 12, oxygen 16), and so a precise determination of the fraction of spin $>0$ nuclei in a given white dwarf would need to be determined to set bounds on spin-dependent dark matter, something that is beyond the scale of this work. Another scenario for which large dark matter-nucleon cross-sections are not yet excluded and could potentially be probed by white dwarf observations is inelastic dark matter \cite{McCullough:2010ai,Bramante:2016rdh}, provided the dark matter settles to the core of the white dwarf ($i.e.$ thermalizes) within the age of the universe.

Turning to neutron stars, a 1.5 solar mass neutron star has escape speed $\sqrt{2GM/R} \sim \frac{2}{3}$ \cite{Shapiro:1983du} and is supported by neutron degeneracy pressure. The extreme velocities and densities mean we must modify Eq.~\eqref{eq:dCn2} to account for two general relativistic corrections when considering dark matter capture on a neutron star.  First, the amount of dark matter crossing the star's surface will be increased because of an enhancement from the star's gravitational potential. It can be shown~\cite{Goldman:1989nd} that for a dark matter particle with velocity $u$ and impact parameter $b$, if the particle barely grazes the surface of the star, then $C_X \propto b^2 = (2 G M R/u^2) [1-2GM/R]^{-1}$, where the square-bracketed term accounts for the general relativistic enhancement to dark matter crossing the star's surface. Accordingly, the dark matter capture rate (with $m = m_n$, of course) is modified to,
\begin{align}
C_N \rightarrow \frac{C_N} {1-\frac{2GM}{R}},
\label{eq:GRcorr}
\end{align}
to account for general relativity-enhanced capture.\footnote{Technically, the general relativistic effects are most straightforwardly introduced into the differential capture rate $dC_N/dr$, which, upon integration, yield Eq.~(\ref{eq:GRcorr}) plus corrections. Given that we are already making an approximation in assuming straight trajectories through the star, we will neglect these corrections to Eq.~(\ref{eq:GRcorr}).} The second general relativistic correction we need is to account for the gravitational blueshift of the dark matter's initial kinetic energy, in the rest frame of a distant observer. In the absence of general relativistic corrections, the dark matter must lose its initial halo kinetic energy $E_i = \frac 1 2 m_X u^2$ via scattering with the star in order to become gravitationally bound to the star. However, from the rest frame of a distant observer, this initial kinetic energy will be enhanced by a factor $\chi = [1-(1-2GM/R)^{1/2}]$\,  under the influence of the star's gravitational potential. This can be accounted for by making the substitution in Eq.~\eqref{eq:uNrel}\,
\begin{align}
v_{\rm esc} \rightarrow  \sqrt{2 \chi}.
\end{align}
In practice, the gravitational and kinetic energy blueshift effects alter the dark matter capture rate in neutron stars by less than a factor of two.

Given the degeneracy of the neutrons that the dark matter must collide with, one may worry that Pauli blocking also comes into play when deriving the capture rate.  Specifically, in order to scatter with the constituents of a neutron star, dark matter must excite them to momenta larger than their Fermi momentum, typically $p_{\rm F,NS } \sim 0.1~ {\rm GeV}$~\cite{Goldman:1989nd}. However, as the incoming dark matter has been accelerated to semi-relativistic speeds in the gravitational well of the neutron star, this requirement is easily satisfied provided the dark matter is heavy. Plugging in numbers, in the limit $m_X \gg m_n$ the average momentum exchanged in any scatter is $Q \sim \sqrt 2\, m_n\, v_{esc} \sim 0.7\, \gev \gg p_{F,NS}$; see $e.g.$ \cite{Bramante:2013hn,Bertoni:2013bsa} for more discussion.

\begin{figure}[h!]
\includegraphics[scale=.75]{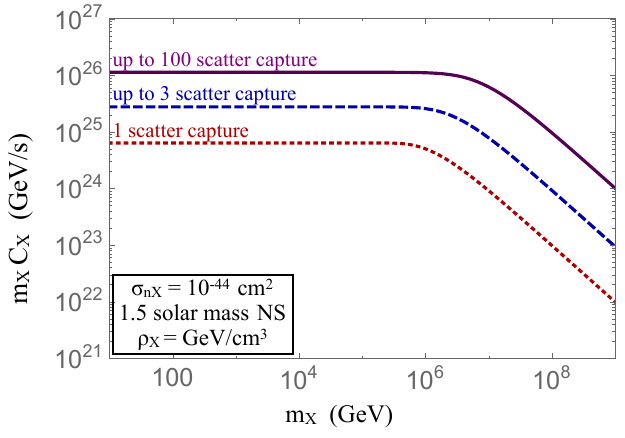}
\includegraphics[scale=.75]{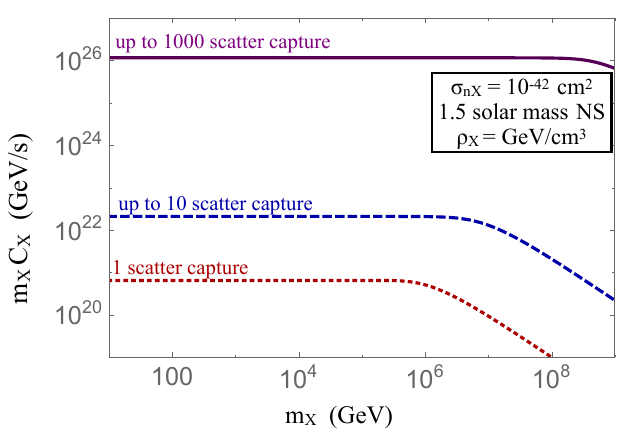}
\caption{Mass capture rate of dark matter on neutron star, for a per-nucleon scattering cross-section of $\sigma_{nX} = 10^{-44}$ and $10^{-42}$ cm$^2$. A constant density, $1.5$ solar mass neutron star composed of neutrons, with $R=10$ km, in a background dark matter density of $\rho_{X} =  {\rm GeV/cm^3}$ with halo velocity dispersion $\bar{v} \simeq 220~{\rm km/s}$ is assumed. Note that the dark matter mass where the mass capture rate shifts from $m_X\, C_X \propto \text{const}$ to $m_X\, C_X \propto 1/m_X$ shifts to higher values as we include more scatters}
\label{fig:nsmc}
\end{figure}

In Fig.~\ref{fig:nsmc} we show the mass capture rate of dark matter on a neutron star for a range of dark matter masses and a dark matter-nucleon cross-sections where $\tau \gtrsim 1$. Figure~\ref{fig:nsmc} has all of the same qualitative features as Fig.~\ref{fig:wdmc}: the mass capture rate increases dramatically once multiple scatters are included, and exhibits a $1/m_X$ dependence in the large $m_X$ limit. However, comparing Figs. \ref{fig:wdmc} and \ref{fig:nsmc}, it is evident that multiscatter capture is relevant for white dwarfs when $\sigma_{nX} \sim 10^{-40}~{\rm cm^2}$, while multiscatter capture on neutron stars becomes important for $\sigma_{nX} \sim 10^{-45}~{\rm cm^2}$. Because the latter cross-section is closer to the cross-section presently probed by direct detection experiments \cite{Aprile:2012nq,Akerib:2016vxi}, we will focus on neutron star probes of dark matter in the next section. While our focus here will be on dark matter which annihilates inside and thereby heats neutron stars, there are many other ways multiscatter stellar capture could be used to probe dark matter, including neutron star implosions \cite{Goldman:1989nd,Kouvaris:2010jy,deLavallaz:2010wp,Kouvaris:2011fi,McDermott:2011jp,Guver:2012ba,Bramante:2013hn,Bell:2013xk,Bramante:2013nma,Bramante:2014zca,Kurita:2015vga,Bramante:2015dfa,Bramante:2016mzo}, monopole-induced nucleon decay \cite{Dimopoulos:1982cz,Kolb:1982si}, white dwarf heating \cite{Hooper:2010es,McCullough:2010ai,Hurst:2014uda}, Type Ia supernova ignition \cite{Bramante:2015cua,Graham:2015apa}, neutrino signatures of superheavy dark matter \cite{Crotty:2002mv,Albuquerque:2002bj}, and dark matter-powered stars \cite{Moskalenko:2007ak,Spolyar:2007qv,Fairbairn:2007bn,Iocco:2008xb}.

\section{Probing heavy dark matter with old neutron stars}
\label{sec:results}

Dark matter that is captured in neutron stars may annihilate to Standard Model particles, thereby heating and increasing the apparent luminosity of old neutron stars. Consequently, the temperature of old neutron stars can be used to probe the dark matter-nucleon cross-section, provided that one bounds or measures the temperature of old stars in regions of sufficiently high dark matter density. Because it harbors a high density of dark matter, the galactic center is an obvious target \cite{Kouvaris:2007ay,Bertone:2007ae,deLavallaz:2010wp,Kouvaris:2010vv}. While old neutron stars at the galactic center are being vigorously sought by the current generation of radio telescopes \cite{Wharton:2011dv,Dexter:2013xga}, to date none have been found, although they are expected to be within reach of next generation radio telescopes like FAST and SKA \cite{FASTSKA}. Here we determine the {\em potential} bounds on dark matter annihilating to SM particles in old neutron stars in the galactic center. Prior work \cite{deLavallaz:2010wp,Kouvaris:2010vv} has explored this bound on dark matter using single scatter capture. This document extends these bounds to higher masses using multiple scatter capture, assuming that DM annihilates to Standard Model particles, and that an old, colder neutron star is resolved in the galactic center at some time in the future. 

The process by which dark matter heats neutron stars involves several steps. First, each captured dark matter particle must thermalize with the host neutron star through successive scatters off neutrons. This thermalization process is complicated by the fact that dark matter momentum will drop after each scatter, and eventually the momentum exchanged between dark matter and the neutrons becomes small enough that Pauli blocking can no longer be ignored. A full calculation of thermalization within neutron stars incorporating Pauli blocking was performed in Ref.~\cite{Bertoni:2013bsa} and showed that the time to thermalize is much less than the age of the neutron star. As one example, for $m_X > 100~{\rm GeV}$ dark matter with a cross-section $\sigma_{nX} > 10^{-48}~{\rm cm^2}$ (well below the values where multscatter becomes important), thermalization occurs in less than a thousand years. Once thermalized, the dark matter settles into a spherical volume $V_{th}$ within the star. Approximating the neutron star as having a constant density core $\rho_{NS}$, $V_{th}$ can be related to the star's temperature $T$ by $V_{th} = \frac{4}{3} \pi r_{th}^3$, $r_{th} = \sqrt{9 T / 4 \pi G \rho_{NS} m_X}$ (see $e.g.$ \cite{Bramante:2015cua}) within the star.

The next step is to understand how $N_X(t)$ -- the number of dark matter particles residing in $V_{th}$ -- evolves with time. Assuming the thermalization time is rapid compared to other timescales, the number of dark matter particles increases as new particles are captured, and decreases as pairs of dark matter particles meet and annihilate. This can be phrased as a simple differential equation for $N_X(t)$~\cite{Bramante:2013nma}, with solution:
\begin{align}
N_X(t)  = \sqrt{\frac{C_{X} V_{th}}{\left\langle \sigma_a v \right\rangle}} {\rm ~tanh} \left[ \sqrt{\frac{C_X \left\langle \sigma_a v \right\rangle  }{V_{th}}}t\right],
\end{align} 
where $C_{X}$ is the net capture rate, $t$ is time over which collection has occurred, and $\left\langle \sigma_a v \right\rangle$ is the thermally-averaged self-annihilation cross-section of the dark matter (DM DM $\to$ SM fields).  Once $t > \sqrt{\frac{V_{th}}{C_X \left\langle \sigma_a v \right\rangle  }}$, the dark matter population plateaus, and there is an equilibrium between the rate at which dark matter is annihilated and the rate at which it is captured.  Assuming all dark matter passing through a neutron star is captured (which implies the longest equilibration time), this equilibration time is \cite{Kouvaris:2010vv}:
\begin{align}
t_{\rm eq} \simeq 10^{4}~{\rm yrs} \left( \frac{10^2~{\rm GeV}}{m_X} \right)^{1/4} \left( \frac{10^3~{\rm GeV/cm^3}}{\rho_{X}} \right)^{1/2}  \left( \frac{T_{NS}}{3 \times 10^4~{\rm K}} \right)^{3/4}\left( \frac{10^{-45}~{\rm cm^3/s}}{\left\langle \sigma_a v \right\rangle} \right)^{1/2},
\label{eq:eqtime}
\end{align}
where $T_{NS}$ is the temperature of the neutron star, and this equilibration time assumes that all DM passing through a $R=10$ km, $1.5~$M$_{\odot}$ neutron star with central density $\rho_{NS} \sim 4*10^{14}~{\rm g/cm^3}$ is captured. The temperatures for the oldest observed neutron stars (age $> 100$ million years) are projected to be $T \ll 2 \times 10^4\, K$~\cite{deLavallaz:2010wp}. Plugging this temperature into Eq.~(\ref{eq:eqtime}) and assuming our local dark matter density $\rho_{X} = 0.3\, \gev/{\rm cm}^3$, we find the equilibration time is $t_{eq} \le 10$ million years for $100\, \gev$ dark matter with annihilation cross sections of $\left\langle \sigma_a v \right\rangle \gtrsim 10^{-48}~{\rm cm^3/s}$. This value is already far less than the age of the oldest neutron stars, and increasing the dark matter mass, density or annihiliation cross section leads to even shorter times; for a benchmark point closer to our region of interest, PeV dark matter in the galactic center ($\rho_{X} = 10^3\, \gev/{\rm cm}^3$) will equilibrate in a $2 \times 10^4\, K$ neutron star in as little as 1000 years if $\langle \sigma_a\, v\rangle = 10^{-45}~{\rm cm}^3/s$.  Because this dark matter self-annihilation cross-section is already quite small, hereafter we assume the dark matter annihilation rate rapidly reaches equilibrium with the capture rate.

\begin{figure}[h!]
\includegraphics[scale=.95]{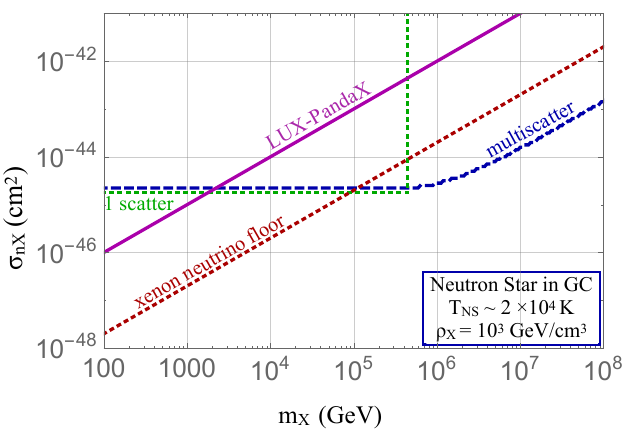}
\caption{Potential sensitivity to dark matter from annihilation to SM particles, heating a $1.5~M_\odot$ neutron star in the galactic center ($\rho_X = 10^3~{\rm GeV/cm^3}$, about 10 parsecs from the Galactic Center) to a local surface temperature of $ \sim 2 \times 10^4$ K, along with interpolations of the current LUX bounds and the neutrino floor (one atmospheric neutrino event on xenon \cite{Billard:2013qya}) for comparison. Here the parameters of the surrounding dark matter density and neutron star temperature have been chosen conservatively; observation of a colder neutron star, or a larger dark matter density would both deepen sensitivity. The curve labeled ``1 scatter" uses Eq.~\eqref{eq:singlescatter} to set the bound, while the multiscatter curve uses the multiscatter formulae derived in this document. Note that multiscatter capture allows for heavier dark matter to be discovered or bounded, for cross-sections below the direct detection neutrino floor.}
\label{fig:ngc}
\end{figure}

Within the parameter space where thermalization and equilibration times are short compared to the typical neutron star lifetime, the annihilation rate is equivalent to the capture rate, and the rate of energy release is simply the mass capture rate $m_X\, C_{X}$. We can define an effective neutron star temperature arising from dark matter annihilations by equating the energy release rate to the apparent luminosity,\footnote{This implicitly assumes that the energy of all DM annihilation products go to heating. It can be verified that the scattering length for neutrinos (and all more strongly coupled Standard Model particles) is much less than the neutron star radius. The exact way the temperature will rise requires knowledge of the equation of state of the star, which is beyond the scope of this paper, but would be an interesting topic for future research.}
\begin{align}
m_X\, C_{X}  = 4 \pi \sigma_0 R^2 T_{NS}^4 \left(1-\frac{2\,G M}{R} \right),
\label{eq:money}
\end{align}
where $\sigma_0 = \pi^2/60$ is the Stefan-Boltzmann constant, and the parenthetical term accounts for enhanced capture of dark matter onto the neutron star.  Read left to right, Eq.~(\ref{eq:money}) defines a minimum temperature for an old neutron star (provided our assumptions of thermalization and equilibration) for a given dark matter mass, density, and capture cross section. Read right to left, Eq.~(\ref{eq:money}) forms a bound. Specifically, if an old neutron star is observed to have surface temperature $T_{NS}$ (as would be seen at the surface of the star), Eq.~(\ref{eq:money}) dictates what regions of $\rho_{X}, m_X$ and $\sigma$ are allowed and which regions would overheat the observed neutron star. Plugging Eq.~\eqref{eq:CNsum} into Eq.~\eqref{eq:money}, we can reframe the expression as
\begin{align}
\sum_{N} p_N(\tau)\left( 1 - \left(1 + \frac{2 A^2_N\, \bar v^2}{3\, v^2_{esc}} \right)\, e^{-A^2_N} \right) = \text{const}\,\frac{T_{NS}^4}{\rho_{X}},
\label{eq:sump}
\end{align}
where the constant on the right hand side is a combination of $G$, $\sigma$, $\bar v$ and the mass and size of the neutron star. The sum over $N$ makes this formula a bit opaque, however we know from Sec. \ref{sec:simpeq} that the left hand side of Eq.~(\ref{eq:sump}) is roughly linear in the dark matter-nucleon cross section $\sigma$ and is either independent of the dark matter mass or $\propto 1/m_X$ depending on whether the dark matter is lighter or heavier than a PeV. Solving Eq.~(\ref{eq:sump}) for $\sigma$, these two regions translate into bounds that are $\sigma \propto \text{const}$ (for $m_X < \text{PeV})$ or $\sigma \propto\, m_X$ (for $m_X > \text{PeV})$. To get a feeling for the type of bound that can be set in this way, in Fig.~\ref{fig:ngc} below we show $\sigma$ could be excluded as a function of $m_X$ should we observe an old neutron star with temperature $T_{NS} \sim 2 \times10^4\, \text K$ in the galactic center ($\rho_{X} = 10^3\, \gev/{\rm cm}^3)$.

There are several interesting features in Figure~\ref{fig:ngc}. First, a shift in the cross section bound around $m_X \sim \text{PeV}$ is evident; this was the mass at which multiple scatter capture becomes relevant, as derived in Section \ref{sec:simpeq}. Second, should a neutron star matching the criteria be found, the DM-nucleon cross section bound it implies would dominate over the existing xenon direct detection bound for all dark matter heavier than $m_X \sim \tev$. Furthermore, while comparing {\em potential} neutron star heating bounds to {\em current} xenon bounds may seem unfair, for $m_X > 0.1\, \text{PeV}$, the cross sections ruled out by neutron star heating are beneath the so-called `neutrino floor' cross section, where direct detection experiments encounter an irreducible background. Given that direct detection experiments are approaching the multi-ton scale and the feasibility of further size increase are far from obvious, observing a cold neutron star may be the best path towards sub-neutrino floor bounds, and further study into how well current and planned telescopes can identify cold neutron stars in environments like the galactic center are warranted~\cite{Bramantetal}. The dependency of the neutron star bound on the temperature of the observed star and the ambient dark matter density where the star is located are clear from the right hand side of Eq. (31), provided one does not deviate too much from the benchmark values of $T_{NS}=2 \times 10^4\, \text K,\, \rho_{X} = 10^3\, \text{GeV/cm}^3$. For example, observing a $T_{NS} \sim  10^4~ {\rm K}$ neutron star in the galactic center would strengthen the bound in Fig.~\ref{fig:ngc} by a factor of $\sim 10$.   For larger temperature or density deviations, the parametrics are not as simple, since the capture rate cannot be increased indefinitely by increasing the DM-nucleus cross section. Specifically, once $\sigma$ reaches the point where all dark matter (at all halo velocities) is captured, further increasing $\sigma$ will not change anything. This `saturation' cross section will depend on the mass of the dark matter.

While a detailed study of the feasibility of constraining neutron stars at various temperatures in the galactic center has not yet been undertaken, we note that observations of $> 10^4\, {\rm K}$ neutron stars within a parsec of the galactic center appear to be within the scope of existing X-ray observatories \cite{Prinz:2015jkd}, and would lead to the strongest bound on the dark matter-neutron cross section for $m_X > {\rm PeV}$.

\section{Conclusions} 
\label{sec:conclusions}

The existence of dark matter has been established by a number of cosmological and astrophysical observations. It is, therefore, one of the most compelling arguments for physics beyond the Standard Model, since there is no candidate for dark matter within the Standard Model. This has inspired vigorous experimental searches for non-gravitational dark matter interactions, including underground detectors looking for dark matter smacking against nuclei, and satellites searching for annihilation of dark matter into Standard Model particles. These searches are most sensitive to dark matter masses up to a few TeV. 

One complementary way to look for heavier dark matter is though its accumulation in stars. Most studies addressing dark matter accumulation in stars have supposed that capture occurs after a single scatter. In this paper we explored  multiscatter capture and found it is particularly relevant for high mass dark matter, which, even for cross-sections below present constraints, will typically scatter multiple times in a neutron star before being captured. We have derived analytical formulae for this process and we have proven that the dark matter-nucleon cross-section bounds obtained at large dark matter masses will have the same parametric dependence as xenon direct detection experiments. Note that while the $\sigma \propto m_X$ scaling at high masses for direct detection experiments is a result of decreased local dark matter number density at high masses ($n_X \sim \rho_X / m_X$), the same parametric dependence that arises for heavy dark matter capture in compact stars results from needing more scattering events to capture higher mass dark matter, as explained in Section \ref{sec:simpeq}.

We have used the resulting formalism to point out bounds on heavy dark matter, which could be obtained through thermal observation of old neutron stars in the galactic center. The resulting bounds are stronger at high dark matter masses, than the reach of next generation direct detection experiments. For $m_X  \gtrsim 100$ TeV the cross-section bound on dark matter that annihilates to Standard Model particles from a $T\sim 10^4$ K neutron star near the galactic center, lies below xenon direct detection cross-sections at which atmospheric neutrinos will begin to provide a substantial background, known as the xenon direct detection neutrino floor. There are additional applications of multiscatter capture, some of which are listed at the end of Section \ref{sec:detail}, which we leave to future work.

{\em Note added:} Following a comment \cite{Ilie:2020vec}, we have corrected typos and an error in our numerical code, which used a white dwarf radius of $10^5$ km instead of $10^4$ km in Figure \ref{fig:wdmc}. Additionally, we have clarified Eq.~\eqref{eq:money} and rounded the neutron star temperature to $\sim 2 \times 10^4~{\rm K}$ in Figure \ref{fig:ngc}, instead of rounding to $\sim 3 \times 10^4~{\rm K}$.  We note that $\sim 2 \times 10^4~{\rm K}$ is an estimate, and likely an underestimate of the neutron star temperature for the parameters given in Figure \ref{fig:ngc}. Factors that will tend to increase the neutron star temperature include curved capture trajectories and a nonuniform neutron star density, as discussed in Section \ref{sec:detail} and recently in \cite{Bell:2020jou}, along with dark kinetic heating \cite{Bramantetal} and crust capture \cite{Acevedo:2019agu}.

\section*{Acknowledgments}
\label{sec:ack}

We thank Masha Baryakhtar, Matthew McCullough, and Nirmal Raj for useful discussions and we thank the authors of \cite{Ilie:2020vec} for reviewing the work presented here. This work was partially supported by the National Science Foundation under Grants No. PHY-1417118 and No. PHY-1520966. Research at Perimeter Institute is supported by the Government of Canada through Industry Canada and by the Province of Ontario through the Ministry of Economic Development \& Innovation.

\appendix
\section{Capture in the optically thin limit}
\label{app:single}
It is useful to summarize the derivation of dark matter capture \cite{Gould:1987ir} on stars:
\begin{itemize}
\item[A.] Far enough away from the star, dark matter particles in the galactic halo have speeds that are Boltzmann distributed. Half the particles will be moving towards the star, namely those with headings $-\pi/2 < \theta < \pi/2$, where $\theta$ is the angle between each particle's velocity and a vector pointing at the star center. The total flux of dark matter is defined as $\mathcal{F}$.
\item[B.] As it traverses the stars gravitational well, the dark matter moves faster in the star's gravitational potential, but conservation of angular momentum implies that its angular momentum with respect to the star remains fixed. Therefore given $\theta$ and the particle's initial speed ($i.e.$ altogether the particle's initial velocity), we can determine whether it has an angular momentum small enough that it will intersect a spherical mass shell at radius $r$ from the center of the star.
\item[C.] The probability that dark matter scatters and is captured while transiting a mass shell of thickness $dr$, depends on the density of scattering sites $n(r)$, the initial dark matter velocity $\vec{u}$, and the dark matter's cross-section with stellar constituents, $\sigma$. Integrating the Boltzmann distributed flux and the probability for capture over $0< u < \infty$ for each stellar mass shell, and integrating mass shells over $0<r<R$, determines the total capture rate. (In the case of multiscatter capture covered in Section \ref{sec:detail}, it is convenient to instead simply consider all dark matter that intersects the star at radius $R$, and then integrate over paths through the star, calculating the multiscatter probability along each path.)
\end{itemize}

We assume dark matter particles surrounding the star will have velocities that follow a Maxwell-Boltzmann distribution. The number density of dark matter particles with velocities ranging from $u$ to $u + du$ is
\begin{align}
f(u)du = 3\sqrt{\frac{6}{\pi}}~ \frac{n_X u^2}{ \bar{v}^3} ~ {\rm Exp} \left[ -\frac{3 u^2}{2 \bar{v}^2}\right]~du,
\label{eq:mboltz}
\end{align}
where $n_X$ is the number density and $\bar{v}$ the average velocity of the dark matter particles. Here $f(u)~ du$ gives the distribution of dark matter velocities far from the gravitational well of the star; nearer to the star each dark matter particle will have a total velocity given by $w^2 = u^2 + v^2(r)$, where $v(r)$ is the escape velocity from the star at radius $r$.

It is useful to at first consider the flux of dark matter particles across a spherical surface large enough that the star's gravitational potential can be neglected. The angle at which dark matter intersects the large surface will increase or diminish its flux across this spherical surface; to account for this, we incorporate a factor of $\vec{u} \cdot \hat{R_{\rm a}} = u ~{\rm cos}~ \theta$, where $\theta$ is the angle between the DM velocity vector $\vec{u}$ and a unit vector $\hat{R_{\rm a}}$ normal to the large surface. Then the flux of dark matter particles towards the star, through an infinitesimal area element, is obtained by integrating the product of $u~ {\rm cos}~ \theta$ and Eq.~\eqref{eq:mboltz} over the range $0 <d ({\rm cos}~\theta) < 1$, and including a factor of $1/2$ to effectively reject the outgoing DM flux,
\begin{align}
d F &=
\frac{1}{2} ~ f(u) u~  du ~ {\rm cos}~ \theta~ d ({\rm cos}~\theta)= 
 \frac{1}{4} ~ f(u) u~ du ~d ({\rm cos^2~}\theta).
\label{eq:influxatinf}
\end{align}

This leads directly to an expression for the flux of dark matter entering a region of size $R_{\rm a}$, which is large enough to ignore the star's gravitational potential,
\begin{align}
d \mathcal{F} &= 4\pi R_{\rm a}^2 ~dF = \pi R_{\rm a}^2~  f(u) u~ du ~d ({\rm cos^2~}\theta).
\label{eq:influxoverR}
\end{align}
To incorporate the star's gravitational potential into the capture rate, we must consider what the dark matter flux will be into a spherical shell of radius $r$, which is the radius of the star or smaller. We define $\alpha$ as the angle between the dark matter particle's velocity vector $\vec{w}$ and the unit normal vector $\hat{r}$ on this small spherical shell. The dark matter's dimensionless angular momentum is
\begin{align}
J \equiv u R_{\rm a} ~ {\rm sin} ~\theta = w r ~{\rm sin} ~\alpha,
\label{eq:Jdef}
\end{align}
where the last equality of Eq.~\eqref{eq:Jdef} follows from angular momentum conservation. As noted previously, $w^2 = u^2 + v^2(r)$, and $v(r)$ is the escape velocity at radius $r$. The flux can now be recast with $dJ^2 = u^2 R_{\rm a}^2 ~d ({\rm cos^2~}\theta)$,
\begin{align}
d \mathcal{F} = \pi f(u) \frac{du}{u}~ d J^2.
\label{eq:dfJ}
\end{align}

As the dark matter particle transits the star's interior, the probability that it is captured after scattering once can be defined as $g_1(w)$. Then the total probability for capture while traversing an infinitesimal spherical shell of length $dl = dr/{\rm cos} ~\alpha$, is the capture probability times the number of path lengths in $dl$:
\begin{align}
 n(r) \sigma  g_1(w) ~dl,
\label{eq:dlprob}
\end{align}
where we have indicated that the number density $n(r)$ of scattering sites may have radial dependence.\footnote{In the case of multiscatter capture, the probability for capturing a dark matter particle that traverses the star in $N$ scatters is given by $g_{N}(w) p_{N}(\tau)$, where these are defined in Section \ref{sec:detail}.} Using Eq.~\eqref{eq:Jdef} to re-express $dl = dr/ \sqrt{1-(J/rw)^2} $, the total single scatter capture rate can then obtained by multiplying Eqs.~\eqref{eq:dlprob} and \eqref{eq:dfJ}, and integrating over $J$. We apply a theta function to require that the dark matter's angular momentum is small enough that it will intersect a shell of size $r$, $\Theta (rw - J)$. We also multiply by a factor of two to account for dark matter passing through both sides of a spherical shell of size $r$, 
\begin{align}
dC_1 &=  4 \pi n(r) \sigma  g_1(w)~ f(u) \frac{du}{u}~ \int_0^{\infty}dJ~\Theta (rw - J)~J ~dl \nonumber \\
 &=4 \pi n(r) \sigma  g_1(w)~ f(u) \frac{du}{u}~ w^2r^2~dr.
\label{eq:g1dCsingle}
\end{align}

It remains to determine the probability for capture after a single scatter, $g_1(w)$. We define
\begin{align}
\beta_{\pm} \equiv \frac{4m_X m}{(m_X \pm m)^2},
\end{align}
where we remind the reader that $m$ is the mass of the stellar constituent with which the DM scatters. A kinematic analysis shows that, in the star's rest frame, the fraction of DM energy lost in a single scatter is evenly distributed over the interval $0 < \Delta E/E_0 < \beta_+$. For single scatter capture, the required fraction of DM kinetic energy loss is $u^2/w^2$, which is the ratio of DM's kinetic energy far away, versus inside the star. To define $g_1(w)$, we use the probability for a single scatter to diminish the DM kinetic energy by a fraction $u^2/w^2$,
\begin{align}
\frac{1}{\beta_+} \left(\beta_+ - \frac{u^2}{w^2} \right),
\label{eq:gouldprob}
\end{align}
along with a theta function that enforces dark matter capture after a single scatter,
\begin{align}
\Theta \left(\beta_+ - \frac{u^2}{w^2} \right).
\label{eq:gouldtheta}
\end{align}
Then $g_1(w)$ is the product of Eqs.~\eqref{eq:gouldprob} and \eqref{eq:gouldtheta}. Inserting this into Eq.~\eqref{eq:g1dCsingle}, and integrating over the incoming Boltzmann distribution of DM ($u$), the total capture rate as a function of radius is
\begin{align}
C_{\rm 1} = \pi \sqrt{\frac{96}{\pi}} \frac{n_{\rm X}}{\bar{v}} \int_{0}^{R} dr~ r^2~ n(r) \sigma(r) v^2(r) \left( 1-\frac{1-e^{-A^2(r)}}{A^2(r)} \right),
\label{eq:singlecapturefull}
\end{align}
where we have indicated that the number density of scattering sites $n(r)$, the escape velocity, $v(r)$, the Boltzmann variable $A^2 \equiv  3v^2(r)/2\bar{v}^2 \beta_-$, and the scattering cross-section, as a consequence of form factor suppression at higher velocities, all depend on the radius of the mass shell, $r$. In the limit that we ignore radial dependence, and set $v(r) \simeq v_{~esc}(R)$ Eq.~\eqref{eq:singlescatter} results.

\bibliography{dmulti.bib}

\end{document}